\documentclass[twocolumn,prl,aps,superscriptaddress]{revtex4-1}
\usepackage[latin9]{inputenc}
\usepackage{color}
\usepackage{braket}
\usepackage{amstext}
\usepackage{amssymb}
\usepackage{graphicx}
\usepackage{soul}
\usepackage[unicode=true,pdfusetitle,
 bookmarks=false,
 breaklinks=false,pdfborder={0 0 1},backref=false,colorlinks=false]
 {hyperref}
\hypersetup{
 bookmarksnumbered=false,bookmarksopen=false}

\makeatletter

\@ifundefined{textcolor}{}
{
 \definecolor{BLACK}{gray}{0}
 \definecolor{WHITE}{gray}{1}
 \definecolor{RED}{rgb}{1,0,0}
 \definecolor{GREEN}{rgb}{0,1,0}
 \definecolor{BLUE}{rgb}{0,0,1}
 \definecolor{CYAN}{cmyk}{1,0,0,0}
 \definecolor{MAGENTA}{cmyk}{0,1,0,0}
 \definecolor{YELLOW}{cmyk}{0,0,1,0}
}

\usepackage{amsmath}
\usepackage{graphicx}
\usepackage{amssymb}
\usepackage{txfonts}

\makeatother

\begin{document}
\title{Modulated Continuous Wave Control for Energy-Efficient Electron-Nuclear Spin Coupling}
\author{J. Casanova}
\affiliation{Department of Physical Chemistry, University of the Basque Country UPV/EHU, Apartado 644, 48080 Bilbao, Spain}
\affiliation{IKERBASQUE,  Basque  Foundation  for  Science,  Maria  Diaz  de  Haro  3,  48013  Bilbao,  Spain}
\author{E. Torrontegui}
\affiliation{Instituto de F\'{\i}sica Fundamental IFF-CSIC, Calle Serrano 113b, 28006 Madrid, Spain}
\author {M. B. Plenio}
\affiliation{Institut f\"ur Theoretische Physik and IQST, Albert-Einstein-Allee 11, Universit\"at Ulm, D-89069 Ulm, Germany}
\author{J. J. Garc{\'\i}a-Ripoll}
\affiliation{Instituto de F\'{\i}sica Fundamental IFF-CSIC, Calle Serrano 113b, 28006 Madrid, Spain}
\author{E. Solano}
\affiliation{Department of Physical Chemistry, University of the Basque Country UPV/EHU, Apartado 644, 48080 Bilbao, Spain}
\affiliation{IKERBASQUE,  Basque  Foundation  for  Science,  Maria  Diaz  de  Haro  3,  48013  Bilbao,  Spain}
\affiliation{Department of Physics, Shanghai University, 200444 Shanghai, China}

\begin{abstract}
We develop energy efficient, continuous microwave schemes to couple electron and nuclear spins, using phase or amplitude modulation to bridge their frequency difference. These controls have promising applications in biological systems, where microwave power should be limited, as well as in situations with high Larmor frequencies due to large magnetic fields and nuclear magnetic moments. These include nanoscale NMR where high magnetic fields achieves enhanced thermal nuclear polarisation and larger chemical shifts. Our controls are also suitable for quantum information processors and nuclear polarisation schemes.
\end{abstract}

\maketitle
\paragraph{Introduction.--} Color centers in diamond, such as the nitrogen-vacancy (NV) center
\cite{Doherty13,Dobrovitski13}, have emerged as a solid state system that can detect,
polarise and control individual nuclear spins in their vicinity \cite{Jelezko04,Muller14,London13,Fisher13}.
This ability promises applications that range from quantum information processing and quantum
simulation on small scale quantum registers \cite{Cai13,Taminiau14,Waldherr14,Casanova15,Cramer16,Casanova16,Perlin17}
to nanoscale nuclear magnetic resonance (NMR) \cite{SchmittGS17,BossCZ17,GlennBL18} and other sensing tasks in biological environments \cite{Wu16}. A fundamental question in this field is how to extend the coherence time of  color centers ---insulating them from their fluctuating magnetic environment---, while enabling strong and selective interactions with individual nuclear spins. For the NV center this challenge is met through dynamical decoupling (DD) schemes: continuous \cite{Bermudez11,Cai12,Hirose12,Cai13a} or pulsed microwave sequences \cite{Maudsley86,Uhrig08,Pasini08,Pasini08bis,Wang11,Souza11,Souza12,Casanova15,Wang16,Lang17,Haase17} that can be applied to mitigate the impact of solid state \cite{Ryan10,NaydenovDH11,Pham12} and biological environments \cite{McGuinness11,Cao17}.

In the context of NMR e.g., the presence of strong magnetic fields would be of great benefit as they increase the NMR signal by enhancing the spin polarisation, induce large chemical shifts that encode molecular structure\ \cite{Levitt08}, and aid in the spectral resolution of spins. Moreover, strong magnetic fields lead to longer nuclear spin lifetimes, facilitating quantum information processors and nuclear polarisation schemes.

Nevertheless, experiments with color centers are typically realised in the sub-Tesla magnetic field regime\ \cite{Gurudev07, Neumann10, Robledo11, Sar12, Zhao12, Kolkowitz12, Taminiau12, Liu13, Muller14, Taminiau14, Waldherr14, Cramer16, Hensen15, Lovchinsky16, Abobeih18} due to experimental limitations. The obstacle is the need to bridge the frequency mismatch between the NV center and the target spin in the presence of a high externally applied magnetic field. When using continuous microwaves, the Larmor frequency of the target nucleus determines the Rabi frequency of the microwave control ---the Hartmann-Hahn (HH) condition\ \cite{Hartmann62}---, implying microwave powers that grow with the magnetic field and imposing serious stability requirements on the microwave source. The situation does not improve for pulsed controls: the Larmor frequency determines the frequency at which $\pi$-pulses are applied to the color center, implying very fast and energetic pulses with high-frequency repetition rates. These power requirements also imply significant challenges for their use in biological samples, because a strong microwave heats the organic matter, perturbing its dynamics or even destroying it. In recent work this challenge was identified and addressed~\cite{Casanova18}. However, in microwave power sensitive applications continuous wave may offer advantages as their average energy consumption at the same decoupling and sensing efficiency can be lower than for pulsed schemes~\cite{Cao17}.

In this Letter, we show that there are indeed continuous microwave controls that can bridge the Larmor frequency difference between electronic and nuclear spins. These methods modulate the phase or amplitude of a continuous microwave field. The modulation is taken to have a frequency $\nu\sim \omega_n-\Omega_0$ that provides the difference between the Rabi frequency of the microwave pulse $\Omega_0$ and the frequency of the target nuclear spin $\omega_n$. This technique works even when the microwave field amplitude $\Omega_0$ is insufficient to achieve a HH resonance. As a result, our schemes demand lower peak and average powers to achieve a coherent interaction with a nucleus than all continuous controls based on the HH condition. Furthermore, we demonstrate that, thanks to the periodic modulation scheme, our controls inherit the robustness against control errors that is typical of DD and pulsed methods.

We start by considering the Hamiltonian of an NV electron spin coupled to a set of nuclei. This reads $(\hbar =1)$
\begin{equation}\label{firsteq}
H = DS_z^2 - \gamma_e B_z S_z - \sum_j \gamma_j B_z I_z + S_z \sum_j \vec{A}_j\cdot \vec{I}_j + H_{\rm c},
\end{equation}
with the NV zero-field splitting $D = (2\pi) \times 2.87$ GHz,  a constant magnetic field $B_z$ applied along the NV axis (i.e. the $\hat{z}$ axis), the gyromagnetic constants for the electronic spin $\gamma_e \approx -(2\pi) \times  28.024$ GHz/T and specific nuclei in the environment $\gamma_j$ ---e.g. $^{13}$C nuclei have $\gamma_{^{13} \rm C}  = (2\pi)\times10.705$ MHz/T---. The NV spin operators are $S_z = \ket{1}\bra{1} - \ket{-1}\bra{-1}$ and $S_x =1/\sqrt{2}(\ket{1}\bra{0} + \ket{-1}\bra{0} + {\rm H.c.})$. The hyperfine vector decays according to a dipole-dipole interaction~\cite{Maze08} $\vec{A}_{j}=\frac{\mu_{0}\gamma_{e}\gamma_{n}}{2|\vec{r}_{j}|^{3}}[\hat{z} - 3\frac{(\hat{z}\cdot\vec{r}_{j})\vec{r}_{j}}{|\vec{r}_{j}|^{2}}]$ with the vector $\vec{r}_j$ connecting the NV center
and the $j$th nucleus. The microwave (MW) control Hamiltonian is conveniently written as $H_{\rm c} =\sqrt{2} \Omega S_x \cos{(\omega t -\phi)}$, parametrized by two external controls: the Rabi frequency $\Omega$ and the microwave phase $\phi$, while the MW frequency $\omega$ will be on resonance with one of the NV spin transitions, namely the $|0\rangle \leftrightarrow |1\rangle$ transition~\cite{Supmat}. Hamiltonian~(\ref{firsteq}) should include the dipole-dipole interaction among nuclei. We omit it to simplify the presentation but it will be fully considered in the numerical simulations below.

An external magnetic field and a suitably tuned microwave field effectively reduce the dimensionality of the NV-center, which can be treated as a pseudospin. The new Hamiltonian~\cite{Supmat}
\begin{equation}\label{ford}
    H = -\sum_j \omega_{n,j} \ \hat{\omega}_{n,j} \cdot \vec{I}_j + \frac{\sigma_z }{2}\sum_j \vec{A}_j\cdot\vec{I}_j + \frac{\Omega}{2} (\ket{1} \bra{0} e^{i\phi} + {\rm H.c.} )
\end{equation}
is defined in a rotating frame generated by $H_0 = DS_z^2 - \gamma_e B_z S_z.$ In this frame, the $j$th nuclear spin's resonance frequency $\omega_{n, j} = |\vec{\omega}_{n, j}|$ with $\vec{\omega}_{n, j} = (-\frac{1}{2} A_{x, j}, -\frac{1}{2} A_{y, j}, \omega_{\rm L} -\frac{1}{2} A_{z, j})$ depends on the hyperfine vectors and the nuclear Larmor frequency $\omega_{\rm L} = \gamma_{j} B_z.$ For simplicity we assume a cluster  of $^{13}$C nuclei $\gamma_j = \gamma_{^{13} \rm C}$ $\forall j$ (a common situation in diamond samples) and introduce the normalized vectors $\hat{\omega}_{n, j} = \vec{\omega}_{n, j}/\omega_{n, j}$.  When the magnetic field $B_z$ is large, the resonance frequency of the $j$th nucleus deviates linearly from its Larmor frequency $\omega_{\rm L}$ as a function of the hyperfine vector
\begin{equation}
  \label{resonancefreq}
  \omega_{n, j} \approx \omega_{\rm L}- \frac{1}{2} A_{z, j} \equiv \gamma_{^{13} \rm C} B_z - \frac{1}{2} A_{z, j}.
\end{equation}
The HH condition~\cite{Hartmann62} is a standard procedure to achieve resonant interaction with a nuclear spin (e.g. the $j$th one) in which the Rabi frequency matches the frequency of the target spin $\Omega =\omega_{n,j}= \gamma_{^{13} \rm C} B_z - \frac{1}{2} A_{z, j}$. In high-field environments this implies large Rabi frequencies and microwave powers ---e.g. $B= 1$ T gives $\Omega/(2\pi) \approx 10$ MHz for a $^{13}$C and $42$ MHz for a $^{1}$H nucleus---. Our goal is to lower these requirements with minor changes in the control field.

\begin{figure}[t]
\hspace{-0.45 cm}\includegraphics[width=1.05\columnwidth]{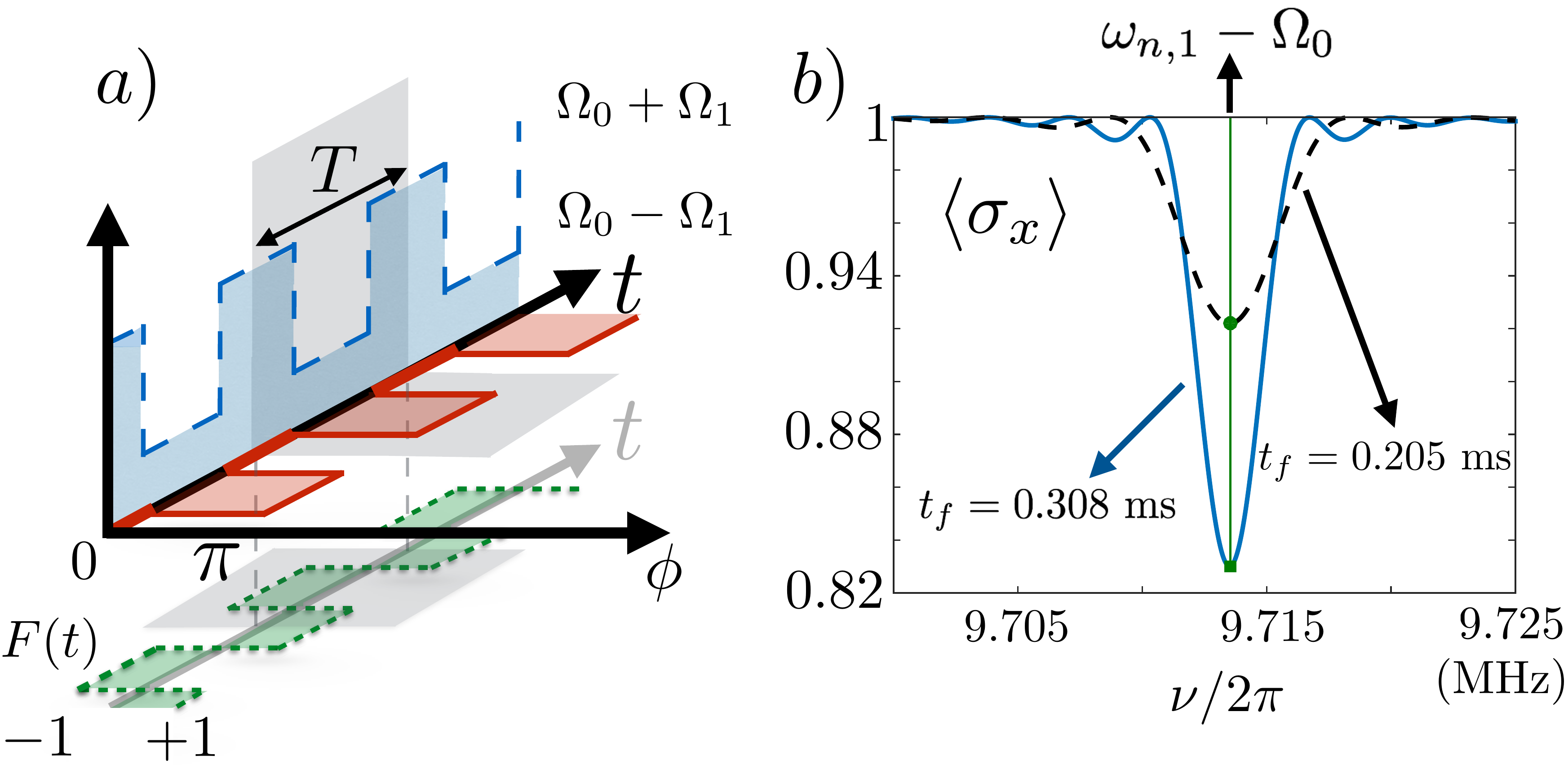}
\caption{a) A discrete phase modulation in time $\phi(t)\in\{0,\pi\}$ (red-solid line) combined with a constant drive, leads to modulations of the Rabi frequency $\Omega(t)=\Omega_0+F(t)\Omega_1\in\{\Omega_0-\Omega_1,\Omega_0+\Omega_1\}$ (blue long-dashed line). Modulation function $F(t)$ (green short-dashed line). b) Harvested signal $\langle\sigma_x\rangle$ vs. phase modulation frequency $\nu=2\pi/T$ for two interrogation times, $t_f=0.205$ ms (black-dashed line) and $t_f = 0.308$ ms (blue-solid line). Signal is maximal when $\nu=\omega_{n,1}-\Omega_0$ spans the difference between the Rabi frequency of the NV and the resonance frequency of the interrogated nuclear spin.}
\label{scheme}
\end{figure}

\paragraph{Phase modulation control scheme.--} We address this challenge and enable NV-nuclear coupling at high magnetic fields by introducing a continuous drive in Eq.~(\ref{firsteq}) described by
\begin{equation}\label{phasemod}
    H_{\rm c} = \sqrt{2} \Omega_0 S_x \cos{(\omega t)}  + \sqrt{2} \Omega_1 S_x \cos{(\omega t - \phi)},
\end{equation}
with a phase $\phi$ that will be switched periodically between the values $0$ and $\pi$. The control in Eq.~(\ref{phasemod}) gives rise to the following Hamiltonian that we will use as the starting point of our simulations~\cite{Supmat}
\begin{equation}\label{transitionH}
H = -\sum_j\omega_{n, j} \hat{\omega}_{n, j} \cdot \vec{I}_j + \frac{\sigma_z}{2} \sum_j \vec{A}_j \cdot\vec{I}_j +\bigg(\frac{\Omega_0 + \Omega_1 e^{i\phi}}{2} \ket{1}\bra{0} + {\rm H.c.} \bigg).
\end{equation}
For the sake of clarity of presentation we consider the phase flips as instantaneous, but stress
that in our numerical simulations the phase flips will take a finite time determined by experimental
limitations. In Eq.~(\ref{transitionH}) the phase flip between $0$ and $\pi$ allows to write the driving
term, i.e. the last term at its right hand side, as $\big[\big(\Omega_0 +F(t) \ \Omega_1\big)/2 \ket{1}\bra{0}
+ {\rm H.c.} \big] $ where the modulation function $F(t)$ takes the values $+1$ (or $-1$) for $\phi=0$
(or $\phi=\pi$). Control of the phase allows for the construction of a modulation function $F(t)$
with period $T$, see Fig.~\ref{scheme}(a), that can be expanded in its Fourier components as $F(t) =
\frac{a_0}{2} + \sum_{n=1}^{\infty}\bigg[ a_n \cos{\bigg(\frac{2n\pi}{T} t\bigg)} + b_n \sin{\bigg(\frac{2n\pi}{T} t\bigg)}\bigg]$ where $a_n = \frac{2}{T}\int_0^T F(t)  \cos{\bigg(\frac{2n\pi}{T} t\bigg)}$, and $b_n =
\frac{2}{T}\int_0^T F(t)  \sin{\bigg(\frac{2n\pi}{T} t\bigg)}$.

To determine the NV-nuclear coupling mechanism and the required resonance condition that result
from the phase control in Eq.~(\ref{phasemod}) we move to a rotating frame with respect to  $-\sum_j\omega_{n, j}
\ \hat{\omega}_{n, j} \cdot \vec{I}_j $ and the driving term $\big[\big(\Omega_0 +F(t) \ \Omega_1\big)/2 \ket{1}\bra{0} + {\rm H.c.} \big] $. For simplicity we select a phase change that produces an
even $F(t)$, i.e. $F(t) = \sum_{n=1}^{\infty} a_n \cos{\big(n\nu t\big)}$ where $\nu = 2\pi/T$ but other constructions including odd components for $F(t)$ are equally possible. This
leads to the Hamiltonian~\cite{Supmat}
\begin{eqnarray}\label{expanded}
    H&&(t)  = \frac{1}{2} \bigg[ \ket{+}\langle -| e^{i\Omega_0 t} e^{i\sum_{n=1}^\infty \frac{a_n \Omega_1}{n\nu} \sin{(n\nu t)}} + {\rm H.c.}\bigg]\cdot\nonumber\\
    &&\sum_j\bigg[ A_{x, j}^\perp I_{x, j} \cos{(\omega_{n, j} t)} + A_{y, j}^\perp I_{y, j} \sin{(\omega_{n, j} t) } + A_{z, j}^{\parallel}I_{z, j} \bigg],
\end{eqnarray}
where $\ket{\pm} = \frac{1}{\sqrt{2}} (\ket{{\rm 1}} \pm \ket{{\rm 0}})$, $A_{x, j}^{\perp} = |\vec{A}_j - (\vec{A}_j \cdot \hat{\omega}_{n, j}) \ \hat{\omega}_{n,j}|$, $A_{y, j}^{\perp} = |\hat{\omega}_{n, j}\times\vec{A}_j|$, $A_{z, j}^{\parallel} = |(\vec{A}_j\cdot\hat{\omega}_{n, j}) \ \hat\omega_{n,j}|$,
$I_{\delta, j} = \vec{I}_j \cdot \hat{\delta}$ with $\hat{\delta} = \hat{x}_j$,  $\hat{y}_j$, or  $\hat{z}_j$,
and $\hat{x}_j = \frac{\vec{A}_j - (\vec{A}_j\cdot\hat{\omega}_{n, j}) \ \hat{\omega}_{n, j}}{A_x^{\perp}}$, $\hat{y}_j = \frac{\hat{\omega}_{n, j}\times\vec{A}_j}{A_y^{\perp}}$, $\hat{z}_j= \frac{ (\vec{A}_j\cdot\hat{\omega}_{n, j}) \ \hat{\omega}_{n, j}}{A_{z, j}^{\parallel}}$. With the aid of the Jacobi-Anger expansion ($e^{iz\sin{(\theta)}}\equiv  \sum_{n=-\infty}^{+\infty} J_n(z) \ e^{i n\theta}$, with $J_n(z)$ the
Bessel function of the first kind) we can rewrite the exponentials in Eq.~(\ref{expanded}) as $e^{i\Omega_0 t} e^{i\sum_{n=1}^\infty \frac{a_n \Omega_1}{n\nu} \sin{(n\nu t)}} =  \prod_{n=1}^\infty \sum_{m=-\infty}^{\infty}  J_m\bigg(\frac{a_n \Omega_1}{n\nu}\bigg)e^{i(\Omega_0  + m  n  \nu) t }$ to find
\begin{equation}\label{branches}
\Omega_0  + m n \nu = \omega_{n, k}
\end{equation}
as the resonance condition for the $k$th nucleus ~\cite{Supmat}. Equation ~(\ref{branches}) implies that,
unlike the HH condition, an NV-nucleus resonance can be achieved for small Rabi frequencies
$\Omega_0$, $\Omega_1$ if we apply a continuous drive interrupted by periodic phase flips at a large frequency
$\nu$. Equation ~(\ref{branches}) exhibits resonances for a wide variety of values $m$ and
$n$ but for small arguments $a_n \Omega_1/(n\nu)$ the interaction strength between the NV and the $k$th
nucleus is largest for $m=n=1$ and $\nu = \omega_{n, k} -\Omega_0$ which yields the effective NV-nucleus
flip-flop Hamiltonian ~\cite{Supmat}
\begin{equation}\label{resonancepulsed}
    H \approx \frac{A_{x,k}^\perp}{2} J_1\bigg(\frac{a_1 \Omega_1}{\nu}\bigg) \bigg[\ket{+}\langle -| I^+_k + \ket{-}\bra{ +} I^-_k  \bigg].
\end{equation}
For the discussion of the energy efficiency it is important to stress at this point that, as we will demonstrate
later, a large value for $\nu$ does not imply large microwave power.

The Hamiltonian (\ref{resonancepulsed}) produces a signal that we will quantify with the electronic expectation
value $\braket{\sigma_x}$, with $\sigma_x = \ket{1}\bra{ 0} + \ket{0} \bra{ 1}$. More specifically, from Eq.~(\ref{resonancepulsed}) one can calculate that the expected signal for a sequence of length $t_f$ is
\begin{equation}\label{signalphasemod}
    \braket{\sigma_x} = \cos^2\bigg[\frac{A^\perp_{x,k} J_1(a_1 \Omega_1/\nu)}{4} t_f\bigg].
\end{equation}
Note that, for a periodic phase-modulated sequence as the one showed in Fig.~\ref{scheme}(a) we have $a_1 = 4/\pi$. Finally, we would like to remark that continuous DD schemes with periodic phase flips have been proposed for extending the NV coherence~\cite{Laraoui11} and to improve DD robustness~\cite{Hirose12}, but their advantages in terms of energy efficiency and nuclear spin control have not been explored to the best of our knowledge.

\paragraph{Amplitude modulation control scheme.--} As an alternative to phase modulation we may also consider
amplitude modulation for achieving energy-efficient electron-nuclear coupling. Let us consider an amplitude modulated continuous
driving field of the form
\begin{equation}\label{ampmod}
    H_{\rm c} =  \sqrt{2} \Omega(t) S_x \cos{(\omega t)}
\end{equation}
with $\Omega(t) = \Omega_0 - \Omega_1 \sin(\nu t)$. Analogously to the previous
section (for more details see~\cite{Supmat}) we find $H(t) = \frac{1}{2} \bigg[ \ket{+}\langle -|  e^{i \Omega_0 t} e^{i\frac{\Omega_1}{\nu}\cos{(\nu t)}}+ {\rm H. c.} \bigg]\cdot \sum_j\bigg[ A_{x, j}^\perp I_x \cos{(\omega_{n, j} t)} + A_{y, j}^\perp I_{y, j} \sin{(\omega_{n, j} t) } + A_{z, j}^{\parallel}I_z \bigg]$.
While we selected an odd amplitude modulation, i.e. a sine-like tailoring for $\Omega_1$, we would like to
stress that other combinations including even modulations are also possible. Again, using the Jacobi-Anger expansion, $e^{iz\cos{(\theta)}}\equiv J_0(z) + 2 \sum_{n=1}^{+\infty} i^n J_n(z)\cos{(n\theta)}$, we find for $\nu = \omega_{n ,k} - \Omega_0$ the following flip-flop Hamiltonian between the NV and the $k$th nucleus~\cite{Supmat} $H \approx \frac{A_{x,k}^\perp}{2} J_1\bigg(\frac{\Omega_1}{\nu} \bigg) \bigg[ i \ket{+} \bra{-} I^+_k - i \ket{-} \bra{+} I^-_k \bigg]$, that leads to  $\braket{\sigma_x} = \cos^2\bigg[\frac{A^\perp_{x,k} J_1(\Omega_1/\nu)}{4} t_f\bigg]$.

\paragraph{Numerical verification.--}
In the following we will analyse the phase-modulated scheme numerically to verify the accuracy of the theoretical
analysis (see~\cite{Supmat} for the analysis of the amplitude modulated scheme which yields similar results). This provides two alternatives which, depending on the specifics of the experimental equipment and the physical
set-up, can be chosen for optimal performance in practice.

To demonstrate the performance of the method, in Fig.~\ref{scheme}(b) we show a spectrum involving an NV center
and a single ${}^{13}$C nucleus such that $\vec{A} \approx (2\pi) \times [-6.71, 11.62, -17.09]$ kHz and
$B =1$ T which results in a nuclear Larmor frequency of $\approx (2\pi) \times 10$ MHz. This hyperfine vector $\vec{A}$ corresponds to a ${}^{13}$C nucleus located in one available position of a diamond lattice. We used two phase-modulated
sequences of different duration (see caption for more details) and show that the obtained signals (yellow-solid
and black-dashed curves) that were numerically computed from Eq.~(\ref{transitionH}) match, firstly,  the position of
the expected resonance $\nu = \omega_{n, 1} - \Omega_0$ for $m=n=1$, see Eq.~(\ref{branches}), and,
secondly, the theoretically calculated depth  (green vertical lines with the circle and square denoting the maximum theoretical depth) for the signal $\braket{\sigma_x}$, see Eq.~(\ref{signalphasemod}), for two different evolution times. Furthermore, in our numerical simulations we did not assume instantaneous
$0$ to $\pi$ phase flips but allowed the phase change to take place in a time interval of length
$t_{\phi_{\rm flip}} \approx 5$ ns, with $\phi$ changing from $0$ to $\pi$ in $20$ discrete steps which is well
within the reach of the time-resolution of modern arbitrary waveform generators~\cite{Naydenovp}. To calculate
the signals in Fig.~\ref{scheme}(b) we used Rabi frequencies $\Omega_0 = \Omega_1 = (2\pi)\times1$ MHz which are one
order of magnitude below the Rabi frequency that would achieve a HH resonance and concomitantly more
energy efficient. Furthermore, our phase modulated scheme allows us to get narrower signals than those obtained with the HH scheme, see Ref.~\cite{Supmat}.

\begin{figure}[t]
    \hspace{-0.3 cm}\includegraphics[width=1.03\columnwidth]{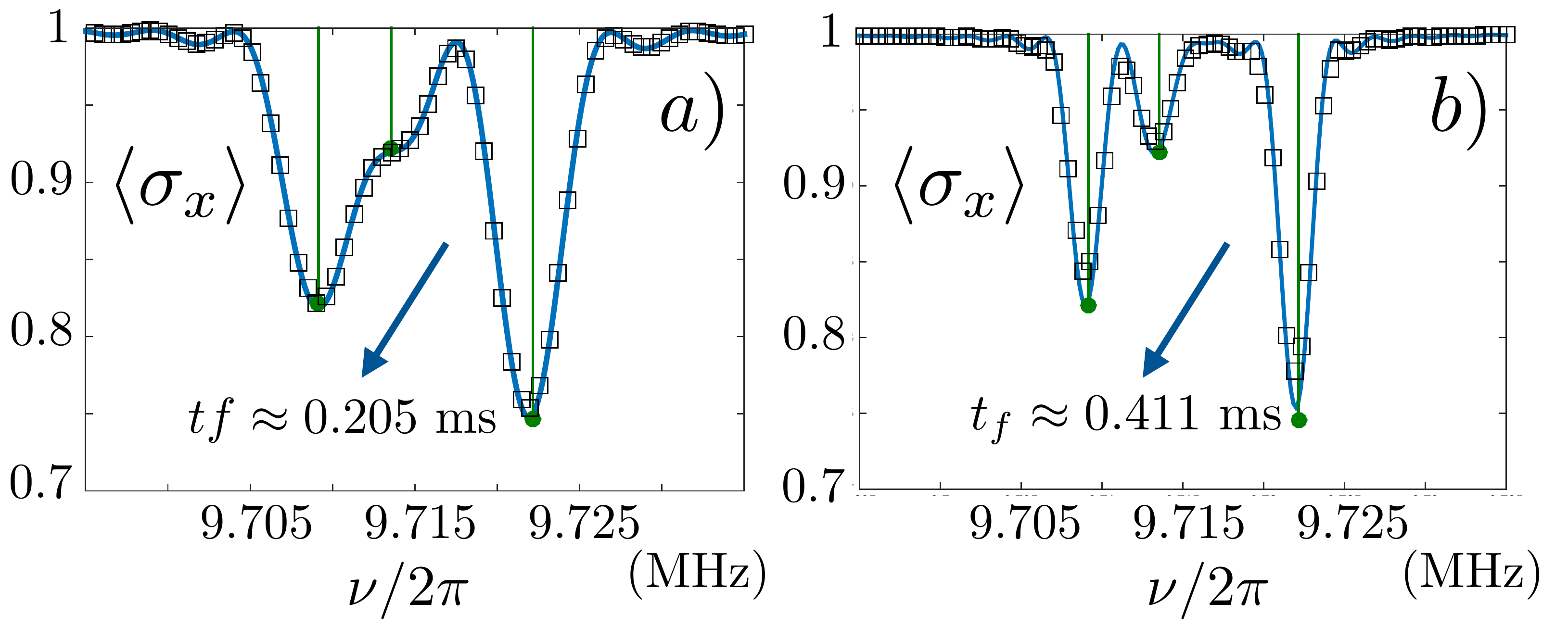}
    \caption{Harvested signal $\braket{\sigma_x}$ as function of $\nu$ for ideal phase-modulated sequences (blue-solid curves) and for a situation that involves fluctuations of the microwave amplitude (black squares). The microwave amplitude fluctuation has been simulated with a OU process, while the black squares have been computed by averaging the results that correspond to 200 experiments. In a) we used $\Omega_0 = \Omega_1 = (2\pi) \times 1$ MHz, while in b) $\Omega_0 = (2\pi) \times 1$ MHz  and  $\Omega_1 = (2\pi) \times 0.5$ MHz. }
\label{threespin}
\end{figure}

\paragraph{Nuclear spin addressing and robustness.--} Our method also offers the possibility of improving
nuclear spin addressing by modifying the value of $\Omega_1$. Hamiltonian~(\ref{resonancepulsed})
shows that the effective NV-nuclear coupling is given by $ A_{x,k}^\perp/2 \ J_1\big(
a_1 \Omega_1/\nu\big) \approx  A_{x,k}^\perp /2 \ \bigg(\frac{a_1 \Omega_1}{2 \nu}\bigg)$. Then,
a lower value $\Omega_1$ implies a longer evolution and better energy selectivity as the rotating wave
approximation over non-resonant terms is more accurate~\cite{Supmat}. In addition, fluctuations on the
microwave control  are also reduced. Note that these are proportional to the Rabi frequency, i.e. $\Omega_{0,1} $ should be replaced by $\Omega_{0,1} [1+ \xi(t)]$ with $\xi(t)$ a fluctuating function. We will show the robustness of our scheme in the face of realistic control errors, see later in Fig.~\ref{threespin}. Furthermore, in Ref.~\cite{Supmat} we study situations with even larger control error conditions, as well as a comparison with the error accumulation process for the case of the HH resonance. 

For the case of NV centers in diamonds with a low Nitrogen concentration, i.e. in ultrapure diamond samples, the main source of decoherence appears as a consequence of the coupling among the NV center and the $^{13}$C nuclei in the lattice~\cite{Maze08, Childress06}. In Fig.~\ref{threespin} we have simulated a system containing
an NV quantum sensor and a three coupled $^{13}$C nuclear spin cluster in a diamond lattice. The hyperfine vectors of the
simulated sample are  $\vec{A}_1
\approx (2\pi) \times [-6.71, 11.62, -17.09]$ kHz, $\vec{A}_2 \approx (2\pi) \times [-8.21, 23.70, -34.30]$ kHz,
and $\vec{A}_3 \approx (2\pi) \times [ 6.76, 19.53, -8.02]$ kHz, such that the resonant frequencies at $B=1$ T,
see Eq.~(\ref{resonancefreq}), are $\omega_{n,1} =(2\pi)\times 10.71$ MHz, $\omega_{n,2} =(2\pi)\times 10.72$ MHz,
and $\omega_{n,3} =(2\pi)\times 10.70$ MHz. These nuclei present  internuclear coupling coefficients $g_{j,l} = (\mu_0/4)(\gamma_{^{13} \rm C}^2/r^3_{j, l}) [1 - 3(n^z_{j, l})^2]$ (with $r_{j, l}$ the distance
between $j$th and $l$th nuclei, and $n^z_{j, l}$ the $z$-projection of the unit vector $\vec{r}_{j, l}/{r}_{j, l}$)  that are $g_{1,2} \approx (2\pi)\times -472$ Hz, $g_{1,3} \approx (2\pi)\times 14.95$ Hz, and $g_{2,3} \approx (2\pi)\times 50.10$ Hz. In Fig.~\ref{threespin}(a) we use a phase-modulated continuous  sequence with $\Omega_0 = \Omega_1 = (2\pi) \times 1$ MHz for a final time $t_f = 0.205$ ms. Here, we have simulated an ideal phase-modulated sequence without microwave control errors (blue-solid line) and a situation involving microwave power fluctuations (black squares). It can be observed that both signals overlap, i.e. the method is noise resilient while, both, the
position of the resonances and the depth of of the signals (green vertical lines) coincide with the theoretical prediction of Eqs.~(\ref{resonancefreq}) and ~(\ref{signalphasemod}), respectively. 
The noise in the microwave field is simulated by averaging 200 runs of a Ornstein-Uhlenbeck (OU) stochastic process~\cite{Uhlenbeck30} with time correlation $\tau = 0.5$ ms and noise amplitude $p = 0.5\%$~\cite{Cohen17}.
In Fig.~\ref{threespin}(a) the three spin resonances cannot be fully resolved, however  in  Fig.~\ref{threespin}(b) we used a different set of control parameters, namely $\Omega_0 = (2\pi) \times 1$ MHz, $\Omega_1 = (2\pi) \times 0.5$ MHz and a longer sequence $t_f = 0.411$ ms. In these conditions the effective NV-coupling has been reduced by a factor of 2 (note that we are also implementing a longer sequence) as the corresponding  Bessel functions are $J_1\big(a_1 [(2\pi)\times 0.5 {\rm MHz}] /\nu\big) \approx \frac{1}{2} J_1\big(a_1 [(2\pi)\times 1  {\rm MHz}]/\nu\big)$, and the three nuclear spins can be clearly resolved. Again, the blue solid line represent the ideal signal while overlapping black squares have been calculated under the same noise conditions than  the previous case.  

Summarizing, from Fig.~\ref{threespin}(b) we can observe how our phase-modulated driving preserves a coherent NV-target nucleus interaction, while eliminating the contributions of the rest of spins in the cluster. Hence, our DD scheme is able to efficiently average-out noisy signals from environmental spins, for more details see Ref.~\cite{Supmat}.

\paragraph{Power consumption.--}
Using phase (or amplitude) modulations leads to a reduction in the microwave field amplitude, i.e. in the peak power, and may be even lead to a reduction in the average power, as compared to other controls using the HH condition. We can quantify the average power reduction comparing the phase modulation scheme with a constant drive based on the HH condition (note that the peak power reduction is obvious as the  largest driving we are using is $(2\pi)\times 1$ MHz which is approximately one order of magnitude smaller than the required driving to hold the HH condition). For that we first identify the times $t^{ph}_f$ and $t^{HH}_f$ for phase modulation and constant HH controls to gather the same signal $\braket{\sigma_x}$. With this information, we can compute the average power or energy flux associated to the microwave control\ \cite{Supmat}.

Let us do the calculation: the protocol with constant amplitude produces a nuclear signal $\braket{\sigma_x}=\cos^2(A_{x,k}^\perp t_f^{HH}/4)$ in a time $t_f^{HH}$. Comparing with Eq. (\ref{signalphasemod}) we find that, for equal signals, the unmodulated protocol implements a  faster interaction in a time $t^{HH}_f=J_1(a_1\Omega_1/\nu)t_f^{ph} < t_f^{ph}$.  During this time, the constant driving scheme requires a Rabi frequency $\bar{\Omega}_0=\omega_{n,k}=\Omega_0 +\nu$ to interact with the $k$-th nucleus with frequency $\omega_{n,k}$. This implies an average energy per cycle $E^{HH}_T \approx \bar{\Omega}_0^2/\nu$. The phase modulated protocol, on the other hand, requires an average energy per cycle $E^{ph}_T\sim (\Omega_0^2+\Omega_1^2)/\nu$. Counting the number of cycles in the respective interaction times $t^{HH}_f$ and $t^{ph}_f$, we obtain the ratio between total powers $E^{HH}/E^{ph}=\nu t_f^{HH}E^{HH}_\nu/(\nu t_f^{ph}E^{ph}_\nu) =(\Omega_0+\nu)^2J_1(a_1\Omega_1/\nu)/(\Omega_0^2+\Omega_1^2)$. We can simplify this formula assuming fast modulation $\nu \gg \Omega_0, \Omega_1$ and approximating  $J_1(a_1\Omega_1/\nu)\sim a_1\Omega_1/(2\nu)$. The result is that the phase modulated protocol demands significantly less energy $E^{HH}/E^{ph}\approx (\Omega_0+\nu)^2\Omega_1a_1/[2\nu(\Omega_0^2+\Omega_1^2)]\gg 1$. For the parameters used in Figs. \ref{threespin}(a) and (b), one finds $E^{HH}/E^{ph}\simeq 3.8$ and $E^{HH}/E^{ph}\simeq 3.0$, respectively, illustrating the efficiency of our method. A similar calculation can be done for the amplitude modulation protocol.

\paragraph{Conclusions.--} We have proposed to use amplitude or phase modulation for coupling electron and nuclear spins at Rabi frequencies well below the Hartmann-Hahn resonance. Our schemes demand lower peak and average power to achieve the same sensitivity. As a consequence, these methods can be employed for sensing, coherent control, and nuclear polarisation with limited accessible power. In particular, they enable the operation of such sensors at high magnetic fields with reduced power. These modulation techniques extend nanoscale NMR techniques to biological systems that are sensitive to heating by microwaves. Moreover, these parametric methods are not specific to the NV center; they can be used to couple different electron spins to proximal nuclear spins, both in solid and molecular samples.

\begin{acknowledgements}
E.S. and J.C. acknowledge financial support from Spanish MINECO/FEDER FIS2015-69983-P, Basque Government IT986-16, as well as from QMiCS (820505) and OpenSuperQ (820363) of the EU Flagship on Quantum Technologies. J.C. acknowledges support by the Juan de la Cierva grant IJCI-2016-29681. E.T. and J.J.G.R. acknowledge support from Spanish MINECO/FEDER Project FIS2015-70856-P, FIS2016-81891-REDT and CAM PRICYT project QUITEMAD+CM S2013-ICE2801. M.B.P. acknowledges support by the ERC Synergy grant BioQ (grant no 319130), the EU project HYPERDIAMOND, the QuantERA project NanoSpin, the BMBF project DiaPol, the state of Baden-W\"urttemberg through bwHPC, and the German Research Foundation (DFG) through grant no INST 40/467-1 FUGG. This material is also based upon work supported by the U.S. Department of Energy, Office of Science, Office of Advance Scientific Computing Research (ASCR), Quantum Algorithms Teams project under field work proposal ERKJ335.
 \end{acknowledgements}

\pagebreak
\widetext
\begin{center}
\textbf{ \large Supplemental Material: \\Modulated Continuous Wave Control for Energy-efficient Electron-nuclear Spin Coupling}
\end{center}
\setcounter{equation}{0}
\setcounter{figure}{0}
\setcounter{table}{0}
\makeatletter
\renewcommand{\theequation}{S\arabic{equation}}
\renewcommand{\thefigure}{S\arabic{figure}}
\renewcommand{\bibnumfmt}[1]{[S#1]}
\renewcommand{\citenumfont}[1]{S#1}

\section{NV-nuclei Hamiltonian with a standard microwave control}
Equation (2) of the main text can be derived from Hamiltonian (1) using an appropriate rotating frame and by selecting only two levels of the NV hyperfine triplet. More specifically, if we start from   
\begin{equation}
H = DS_z^2 - \gamma_e B_z S_z - \sum_j \gamma_j B_z I_z + S_z \sum_j \vec{A}_j\cdot \vec{I} + H_{\rm c},
\end{equation}
where the external microwave control term given by $H_{\rm c} =\sqrt{2} \Omega S_x \cos{(\omega t -\phi)}$, we see that in a rotating frame whit respect to (w.r.t.)  $DS_z^2 - \gamma_e B_z S_z $ the above Hamiltonian becomes 
\begin{equation}
H = - \sum_j \gamma_j B_z I_z +(|1\rangle\langle 1| -  |-1\rangle\langle -1|) \sum_j \vec{A}_j\cdot \vec{I} + \frac{\Omega}{2}\bigg[ |1\rangle\langle 0| e^{i(D+|\gamma_e| B_z)t} + |-1\rangle\langle 0| e^{i(D-|\gamma_e| B_z)t}  + {\rm H.c.}\bigg] (e^{i\omega t} e^{-i\phi}+ e^{-i\omega t} e^{i\phi}),
\end{equation}
with $S_z = |1\rangle\langle 1| -  |-1\rangle\langle -1|$ and $S_x = \frac{1}{\sqrt{2}} \big(|1\rangle\langle 0| + |-1\rangle\langle 0| + {\rm H.c.}\big)$. Now, setting the microwave field frequency $\omega = D+|\gamma_e| B_z$ we can eliminate counter rotating terms in the 
control term leading to the form $\frac{\Omega}{2} (|1\rangle \langle 0| e^{i\phi} + {\rm H.c.} )$ that appears in Hamiltonian (2) of the main text. In addition, in the above Hamiltonian we can also eliminate the term $ |-1\rangle\langle -1| \sum_j \vec{A}_j\cdot \vec{I}$ since the $|-1\rangle$ state of the NV does not enter in the dynamics, as there are no terms in the Hamiltonian that produces transitions to this level. Thus, we can write 
\begin{equation}
|1\rangle\langle 1| \sum_j \vec{A}_j\cdot \vec{I} = \frac{1}{2} (|1\rangle\langle 1| - |0\rangle\langle 0|) \sum_j \vec{A}_j\cdot \vec{I} + \frac{1}{2} (|1\rangle\langle 1| + |0\rangle\langle 0|) \sum_j \vec{A}_j\cdot \vec{I}. 
\end{equation}
Note that as $|-1\rangle$ has been discarded from the dynamics, $|1\rangle\langle 1| - |0\rangle\langle 0| = \sigma_z$ and $|1\rangle\langle 1| + |0\rangle\langle 0| = \mathbb{I}$, so the $|0\rangle$ and $|1\rangle$ hyperfine spin states form a complete basis for the NV spin degrees of freedom.
Adding the term $\mathbb{I} \sum_j \vec{A}_j\cdot \vec{I} $  to the nuclear Larmor frequency we finally obtain
\begin{equation}
 - \sum_j \gamma_j B_z I_z + \mathbb{I}  \sum_j \vec{A}_j\cdot \vec{I}  = -\sum_j \omega_{n,j} \ \hat{\omega}_{n,j} \cdot \vec{I}_j 
\end{equation}
and, consequently, the Hamiltonian (2) in the main text is recovered.

\section{NV-nuclei Hamiltonian with phase modulated microwave control}
The procedure to get Hamiltonian (5) in the main text that includes a phase-modulated control is similar to the one  described in the previous section (i.e. one should move to a rotating frame, eliminate the counter rotating terms and the $|-1\rangle$ level, and re-order the remaining terms in the Hamiltonian) but now with a microwave control term of the form 
\begin{equation}
H_{\rm c} = \sqrt{2} \Omega_0 S_x \cos{(\omega t)}  + \sqrt{2} \Omega_1 S_x \cos{(\omega t - \phi)}
\end{equation}
where the phase $\phi$ will change from $0$ to $\pi$ and viceversa in a time interval $t_{\rm flip}$.
\section{Interacting Hamiltonian and resonance branches}
The Hamiltonian in Eq.~(5) of the main text in a rotating frame w.r.t. 
$H_0(t) = -\sum_j\omega_{n, j} \ \hat{\omega}_{n, j} \cdot \vec{I}_j  + \big[\big(\Omega_0 +F(t) \ \Omega_1\big)/2 |1\rangle\langle0| + {\rm H.c.} \big]$ with $F(t) = \sum_{n=1}^{\infty} a_n \cos{\big(n\nu t\big)}$ reads 
\begin{equation}\label{sup:resonances}
H =e^{i \int_0^t H_0(s) ds}  \frac{\sigma_z}{2} \sum_j \vec{A}_j \cdot\vec{I}_j e^{-i \int_0^t H_0(s) ds} = \frac{1}{2} \bigg[ |+\rangle\langle -| e^{i\Omega_0 t} e^{i\sum_{n=1}^\infty \frac{a_n \Omega_1}{n\nu} \sin{(n\nu t)}} + {\rm H.c.}\bigg]
\sum_j\bigg[ A_{x, j}^\perp I_{x, j} \cos{(\omega_{n, j} t)} + A_{y, j}^\perp I_{y, j} \sin{(\omega_{n, j} t) } + A_{z, j}^{\parallel}I_{z, j} \bigg],
\end{equation}
where we have used the property 
\begin{equation}\label{sup:identity}
e^{i\vec{I}_{j}\cdot\hat{l}\phi}\vec{I}_{j}\cdot\vec{b}\ e^{-i\vec{I}_{j}\cdot\hat{l}\phi}=\vec{I}_{j}\cdot\{[\vec{b}-(\vec{b}\cdot\hat{l})\hat{l}]\cos\phi-\hat{l}\times\vec{b}\sin\phi+(\vec{b}\cdot\hat{l})\hat{l}\}.
\end{equation}  

Now, the strategy to search for resonances is to look for the time independent terms  in Eq.~(\ref{sup:resonances}) that appear by multiplying the exponentials $e^{i\Omega_0 t} e^{i\sum_{n=1}^\infty \frac{a_n \Omega_1}{n\nu} \sin{(n\nu t)}} = \prod_{n=1}^\infty \sum_{m=-\infty}^{\infty}  J_m\bigg(\frac{a_n \Omega_1}{n\nu}\bigg)e^{i(\Omega_0  + m  n  \nu) t }$ with the terms $e^{\pm i \omega_{n,j}t}$ that emerge from the functions $\cos{(\omega_{n, j} t)}$ and $\sin{(\omega_{n, j} t) }$. In this manner, these time independent terms come out when the following condition (this is Eq.~(7) in the main text) applies 
\begin{equation}\label{sup:branches}
\Omega_0  + m n \nu = \omega_{n, k}.
\end{equation}

\section{Resonant Hamiltonians and Nuclear Spin Addressing}
\subsection{Resonant Hamiltonians}
For the case $n=m=1$ and when searching around a certain resonance, for example the one of the $k$th nucleus which corresponds to scan for a set of values of $\nu$ such that $\nu \approx \omega_{n,k} - \Omega_0$,  the Hamiltonian in Eq.~(6) of the main text, or Hamiltonian~(\ref{sup:resonances}) in this Suplemental Material is (note that other cases for arbitrary $n$ and $m$ work in the same manner)
\begin{equation}
H \approx J_1\bigg(\frac{a_1\Omega_1}{\nu}\bigg)/4\bigg[  |+\rangle\langle -| e^{i(\Omega_0  + m  n  \nu) t } + {\rm H.c.} \bigg]\bigg[A_{x,k}^\perp I_{x, k} (e^{i\omega_{n, k}t}+ e^{-i\omega_{n, k}t}) - i A_{y,k}^\perp I_{y, k} (e^{i\omega_{n, k}t} - e^{-i\omega_{n, k}t}) + A_{z,k}^\parallel I_{z, k}\bigg].
\end{equation}
The term involving the $z$-components of the hyperfine field, $A_{z,k}^\parallel$ is eliminated because of the fast oscillatory functions  
$e^{i(\Omega_0  + m  n  \nu) t }$. Imposing the resonant condition  $\nu \approx \omega_{n,k} - \Omega_0$ we  get 
\begin{equation}\label{sup:detuned} 
H \approx \frac{A_{x,k}^\perp}{2} J_1\bigg(\frac{a_1\Omega_1}{\nu}\bigg) \bigg[  |+\rangle\langle -| e^{i\delta t } I^{+}_k + {\rm H.c.} \bigg]
\end{equation}
where $\delta = \Omega_0 + \nu - \omega_{n, k}$. In this manner, when $\delta \gg \frac{A_{x,k}^\perp}{2} J_1\bigg(\frac{a_1\Omega_1}{\nu}\bigg)$ there is no signal as the RWA applies averaging out the Hamiltonian in  Eq.~(\ref{sup:detuned}). In addition, if $\delta = 0$ we get Eq.~(8) of the main text.
\subsection{Nuclear Spin Addressing}
The presence of the $J_1\bigg(\frac{a_1\Omega_1}{\nu}\bigg)$ functions in the effective interactions that emerge from our method allows to induce a better nuclear spin selectivity. For example, if we consider a situation involving two nuclear spins with resonance frequencies $\omega_{n,1}$ and $\omega_{n,2}$, and if we set $\nu$ such that we are on resonance with the first nucleus, i.e. $\nu = \omega_{n,1} - \Omega_0$, the effective Hamiltonian that appears is
\begin{equation}
H \approx \frac{A_{x,1}^\perp}{2} J_1\bigg(\frac{a_1\Omega_1}{\nu}\bigg) \bigg[  |+\rangle\langle -|  I^{+}_1 + {\rm H.c.} \bigg] + \frac{A_{x,2}^\perp}{2} J_1\bigg(\frac{a_1\Omega_1}{\nu}\bigg) \bigg[  |+\rangle\langle -| e^{i(\omega_{n,1} - \omega_{n,2} )t} I^{+}_2 + {\rm H.c.} \bigg].
\end{equation}
Then, if we want to address the first nuclear spin, we need to eliminate the coupling with the second nucleus. The latter can be done thanks to the presence of the function $J_1\bigg(\frac{a_1\Omega_1}{\nu}\bigg)$ 
in the effective NV-nucleus coupling, its value can be easily tuned by delivering a lower microwave intensity, i.e. a small value for $\Omega_1$. This effect can be seen in Fig.~2. of the main text for a coupled nuclear cluster involving three spins.

\begin{figure}[t!]
\hspace{-0.7 cm}\includegraphics[width=0.9\columnwidth]{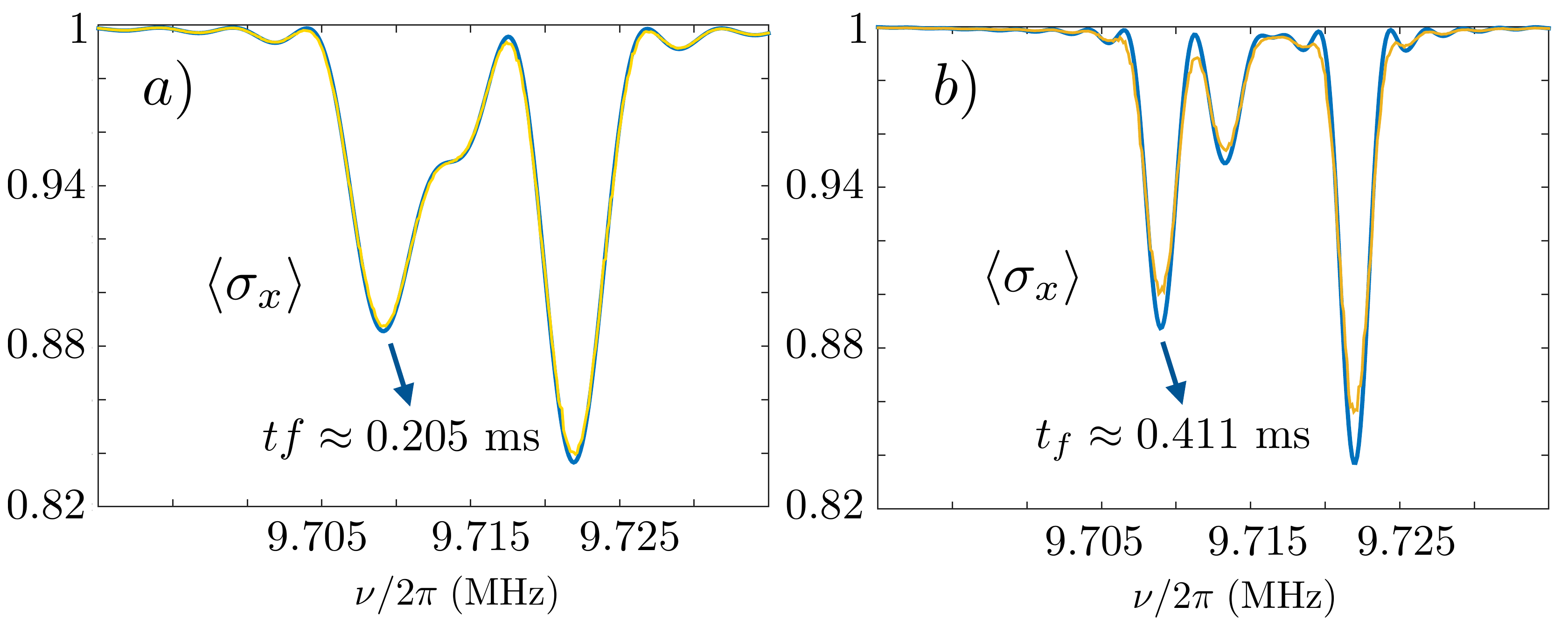}
\caption{Nuclear signal (solid blue and yellow) of a NV coupled to three nuclear spins. The final  time of the evolution is a) $t_f \approx 0.205$ ms and b) $t_f \approx 0.411$ ms. For the simulations we used nuclei with hyperfine vectors such that  $\vec{A}_1 = (2\pi) \times [-6.71, 11.62, -17.09]$ kHz, $\vec{A}_2 = (2\pi) \times [-8.21, 23.70, -34.30]$ kHz, and $\vec{A}_3 = (2\pi) \times [67.68, 195.39, -82.90]$ kHz. The noisy cases (yellow lines) include in the simulations fluctuations of the driving amplitude according to a OU process with 200 runs.}
\label{sup:addressing} 
\end{figure}
\section{The Method with Amplitude Modulated microwave radiation}
In this section we show the performance of the amplitude modulated method for achieving nuclear spin selectivity. More specifically we used the following Hamiltonian
\begin{equation}
H = -\sum_j\omega_{n, j} \hat{\omega}_{n, j} \cdot \vec{I}_j + \frac{\sigma_z}{2} \sum_j \vec{A}_j \cdot\vec{I}_j +\bigg[\frac{\Omega_0 - \Omega_1\sin{(\nu t)}}{2} \bigg] \bigg[|1\rangle\langle0| +  |0\rangle\langle1| \bigg].
\end{equation}
as the starting point of our simulations. 

 In Fig.~\ref{sup:addressing} we have simulated the same nuclear spin cluster used in the main text for two  amplitude-modulated microwave sequences of different duration (see caption for details). Interestingly, although the amplitude-modulated sequences are displayed for the same time than the phase-modulated sequences presented in Fig.~2 of the main text, the former lead to a weaker signal. This is because the resonant Hamiltonian for the amplitude-modulated sequences is 
\begin{equation}
H \approx \frac{A_{x,k}^\perp}{2} J_1\bigg(\frac{\Omega_1}{\nu} \bigg) \bigg[ i |+\rangle \langle-| I^+_k - i |-\rangle \langle+| I^-_k \bigg],
\end{equation}
where for the effective coupling appears the term $J_1\bigg(\frac{\Omega_1}{\nu} \bigg)$ which is smaller than $J_1\bigg(\frac{\Omega_1 a_1}{\nu} \bigg)$ (as $a_1 = 4/\pi > 1$) that it is present in the resonant Hamiltonian for phase-modulated sequences, see Eq.~(8) in the main text.

\section{Full width at the half maximum and the comparison with the HH method}
\begin{figure}[t!]
\hspace{-0.7 cm}\includegraphics[width=0.8\columnwidth]{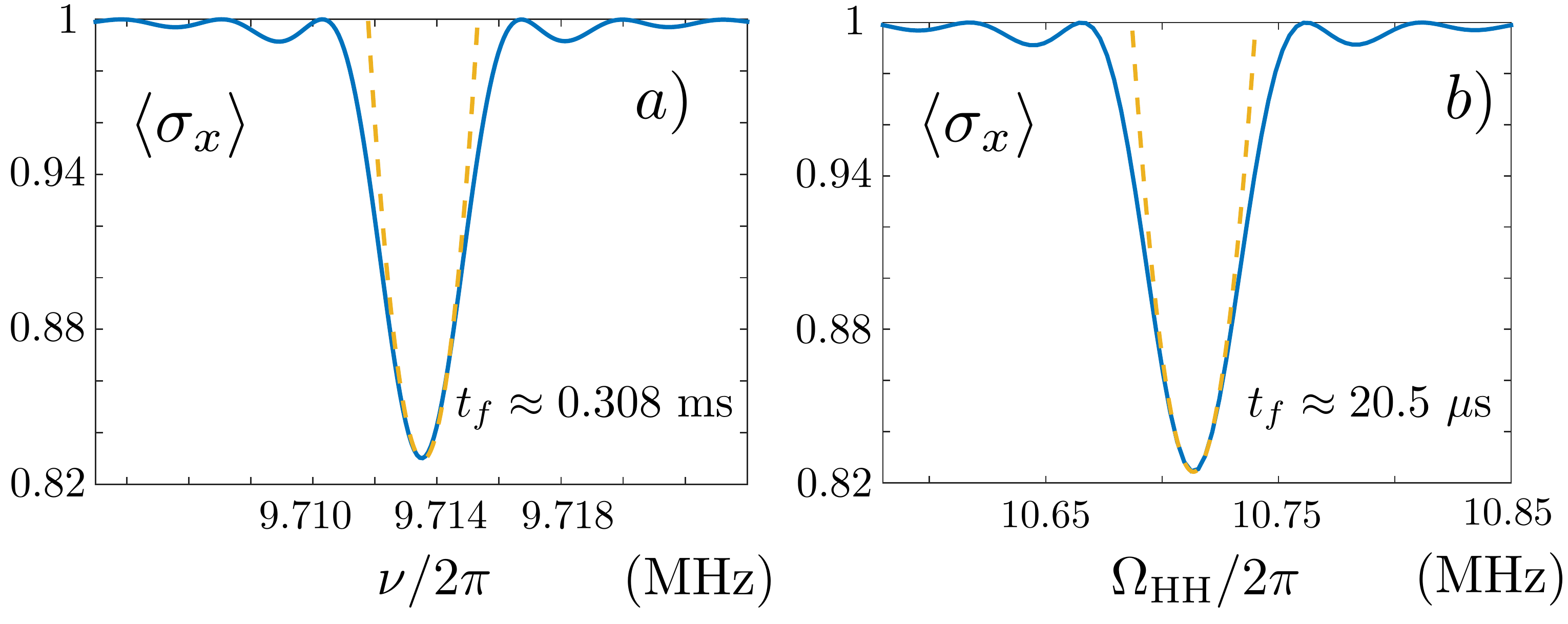}
\caption{A comparison between numerical results for, a), our phase modulated method (solid-blue) and the harmonic approximation (dashed-yellow), and b) the HH method (solid-blue) and the harmonic approximation (dashed-yellow). We can observe, see horizontal axes, how our method leads to narrower signals. We used the same nuclear spin as in Fig.~1 of the main text. This is a nucleus with hyperfine vector $\vec{A} = (2\pi) \times [-6.71, 11.62, -17.09]$. For these values, the
numerical results are indistinguishable from the analytic prediction in Eq. (\ref{signalA}).}
\label{sup:FWHM} 
\end{figure}
In the following, we derive an analytical expression that describes the harvested signal when addressing a single nuclear spin, and its associated full width at the half maximum (FWHM). To this end, we assume that
near one nuclear resonance the system is described by Eq.~(\ref{sup:detuned}). In order to eliminate the explicit time dependency in that Hamiltonian, we move to an interaction 
picture with respect to $H_0=-\delta I_{z,k}/2$. In this representation, the system is governed by $H_I=e^{iH_0t}(H-H_0)e^{-iH_0t}$, explicitly,
\begin{equation}
H_I=\frac{A_{x,k}^{\perp}}{2}J_1\bigg(\frac{a_1\Omega_1}{\nu}\bigg)[|+\rangle\langle-|I_k^{+}+{\rm H.c.}]+\frac{\delta}{2}I_{z,k}.
\end{equation}
 The harvested signal is now easily computed as $\langle\sigma_x\rangle = \mbox{Tr} (e^{-iH_0t_f}e^{-iH_It_f}\rho_0e^{iH_0t_f}e^{iH_It_f}\sigma_x)$.
 Assuming that initially the NV is in the $|+\rangle\langle +|$ state and the nucleus is in a thermal state represented by the 2 $\times$ 2 identity operator, we have
\begin{equation}
\label{signalA}
\langle\sigma_x\rangle = \frac{e^{-i\delta t_f}\bigg[-A^{\perp^2}_{x,k}J_1^2+e^{i\delta t_f}(16\delta^2+3A^{\perp^2}_{x,k}J_1^2)+A^{\perp^2}_{x,k}(1+e^{i\delta t_f})J_1^2\cos \bigg(\frac{t_f}{2}\sqrt{4\delta^2+A^{\perp^2}_{x,k}J_1^2}\bigg)\bigg]}{4(4\delta^2+A^{\perp^2}_{x,k}J_1^2)}, 
\end{equation}
with $J_1\equiv J_1\big(\frac{a_1\Omega_1}{\nu}\big)$. With this equation we can estimate the width of the signal around the resonance point $\delta\approx 0$ in the $\nu\gg\{\Omega_0, \Omega_1\}$ regime. By writing 
$\langle\sigma_x\rangle \approx \mathcal{A}+\mathcal{B}\delta^2$, with $\mathcal{A}$ being the value of $\langle\sigma_x\rangle$  given by Eq. (9) of the main text (i.e. it is the value of  $\langle\sigma_x\rangle$ at a certain nuclear resonance) and expanding a Gaussian to the harmonic approximation $f(\delta)=a e^{\delta^2/(2b^2)}\approx a+\frac{a\delta^2}{2b^2}$, the $\mbox{FWHM}$ for our phase modulated scheme is $\mbox{FWHM}^{ph}$ $\propto b\propto 1/\sqrt{\mathcal{B}}=\frac{16\pi^2 \nu}{A_{x,k}^{\perp}a_1\Omega_1t_f^2}$.
Following a similar derivation, we can estimate the width of the harvested signal for the Harman-Hahn method. This is, one starts from Eq.~(\ref{sup:detuned}) and obtains Eq.~(\ref{signalA}), but replacing in both equations the $J_1\big(\frac{a_1\Omega_1}{\nu}\big)$ coefficient by 1. For this method in the  $\delta\approx 0$ and $\nu\gg\Omega_0$ regime the signal is given
by $\langle\sigma_x\rangle \approx \cos^2(A_{x,k}^{\perp}t_f/4)+\mathcal{B}\delta^2$ with a $\mbox{FWHM}^{HH}$ $\propto 1/\sqrt{\mathcal{B}}=\frac{8\pi^2}{ A_{x,k}^{\perp}t_f^2}$. In Fig.~\ref{sup:FWHM}(a) we show a comparison of the harmonic approximation accuracy (dashed-yellow) with the numerical results of our phase modulated scheme (solid-blue). The same comparison is done for the case of the HH method in Fig.~\ref{sup:FWHM}(b). To achieve the same on-resonance strength ($\mathcal{B}=0$) the required times are $t_f^{HH}=J_1(a_1\Omega_1/\nu)t_f^{ph}$, the phase modulation control scheme produces a narrower signal (i.e. $\mbox{FWHM}^{HH}/\mbox{FWHM}^{ph}\approx 2\nu/(a_1\Omega_1)\gg 1$) compared with the Hartmann-Hahn sensing protocol. In this respect, see horizontal axes in Figs.~\ref{sup:FWHM}(a) and (b).

\section{Larger MW control noise}
\begin{figure}[t!]
\hspace{-0.7 cm}\includegraphics[width=0.75\columnwidth]{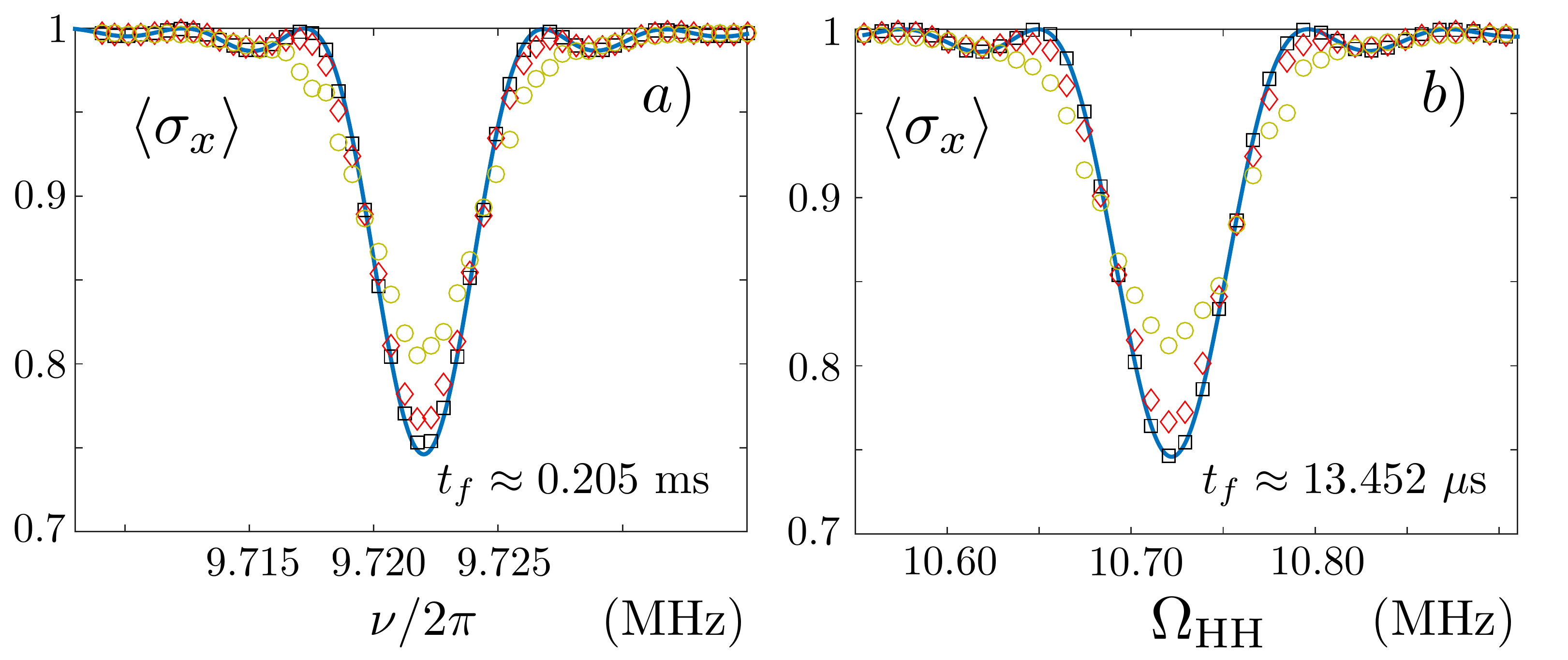}
\caption{Harvested signal for growing noise amplitude. In a) we plot the achieved signal when our method is used (the final time of the simulated phase modulated sequence is $\approx 0.205$ ms). In b) we simulate the case of HH double resonance method that requires less time for the same signal depth ($t_f\approx 13.452 \ \mu$s) and, therefore, the noise influence gets reduced. }
\label{sup:comparisonHH} 
\end{figure}
We have simulated situations with stronger error conditions and made a comparison with the case of the standard HH double resonance scheme. In Fig.~\ref{sup:comparisonHH} we include one of the nuclear resonances of Fig.~2(a) in the main text with growing amplitude noise. More specifically, in Fig.~\ref{sup:comparisonHH}(a) we have the spectrum without MW noise (blue-solid curve) obtained with our phase modulated scheme, and the cases with MW noise modeled with a Ornstein-Uhlenbeck (OU) stochastic process with time correlation $\tau=0.5$ ms and noise amplitude p=0.5 $\%$ (black squares), p=1.0 $\%$ (red diamonds), and p=2.0 $\%$ (green circles). In this figure we can observe a signal decay of $\approx 20 \%$ on the resonance position for the largest amplitude noise  (green circles). In Fig.~\ref{sup:comparisonHH}(b) we have simulated the ideal HH resonance process, i.e. in the absence of any MW noise (blue-solid curve), for a maximum signal depth similar to the one we got in Fig.~\ref{sup:comparisonHH}(a). This is a signal depth of $\approx 0.75$. Note however that, the frequency width in Fig.~\ref{sup:comparisonHH}(a) (i.e. the one achieved with our method) is much narrower than in Fig.~\ref{sup:comparisonHH}(b), see horizontal axes in Figs.~\ref{sup:comparisonHH}(a) and (b). The latter is of great help for single nuclear spin determination and represents another advantage of our introduced phase modulated method. Furthermore, in Fig.~\ref{sup:comparisonHH}(b) we have included the cases of a noisy MW driving for the HH scheme. As in the previous case, the noise is modelled with a OU process with time correlation $\tau=0.5$ ms and noise amplitude p=0.5 $\%$ (black squares), p=5.0 $\%$ (red diamonds), and p=10 $\%$ (green circles). In this manner, we can observe that, as the HH method requires less time than our protocol, the impact of the microwave noise is less severe.

\section{Further details on decoherence and noise elimination}
\begin{figure}[t!]
\hspace{-0.7 cm}\includegraphics[width=0.9\columnwidth]{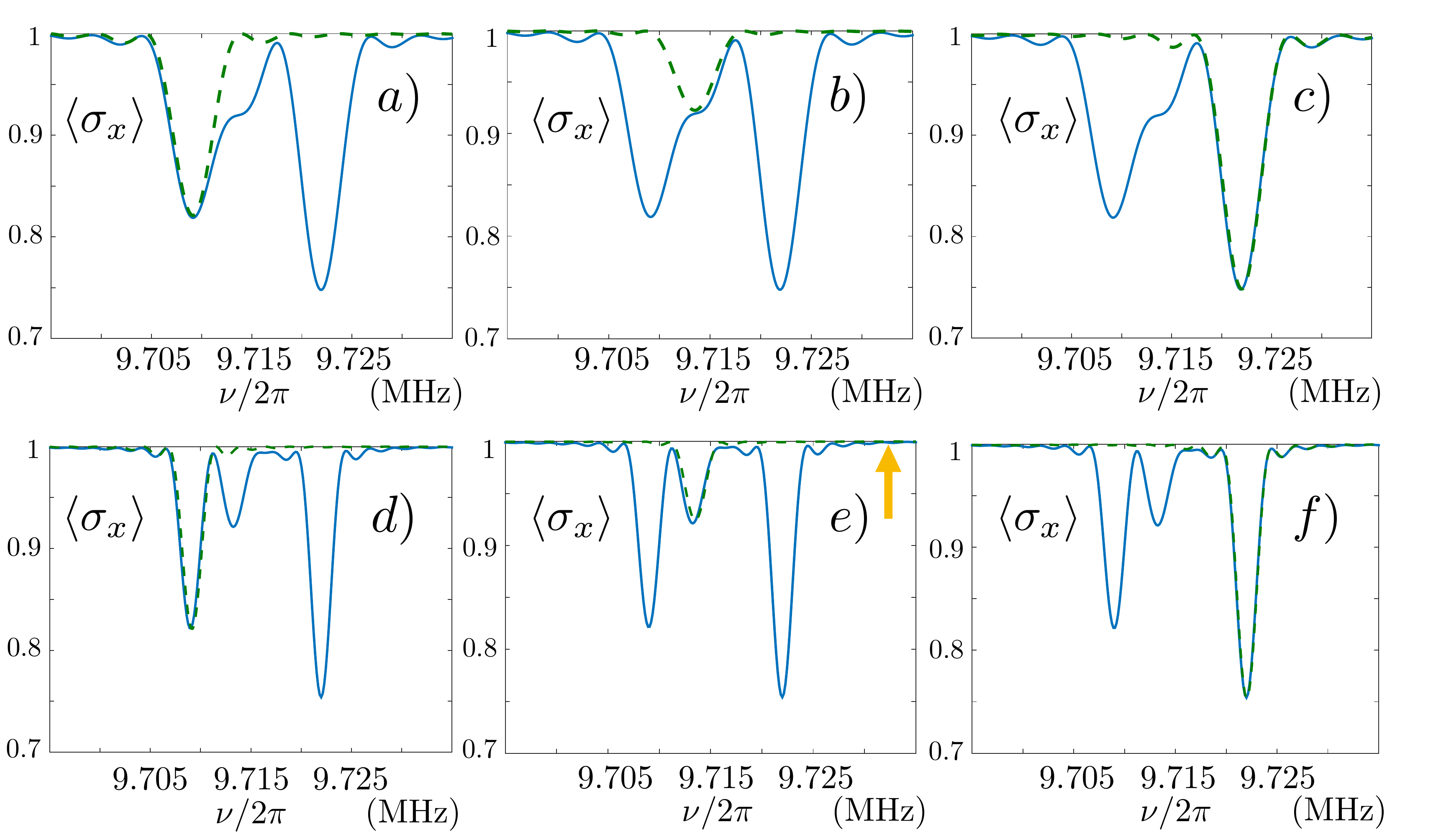}
\caption{Nuclear spin cluster signal (solid-blue) compared with each individual nuclear spin signal (green-dashed) for different values of the control parameters. We used the same nuclear spin cluster that appears in Fig.~2 of the main text.  This is, we include nuclei with the following hyperfine vectors $\vec{A}_1 = (2\pi) \times [-6.71, 11.62, -17.09]$ kHz, $\vec{A}_2 = (2\pi) \times [-8.21, 23.70, -34.30]$ kHz, and $\vec{A}_3 = (2\pi) \times [6.76, 19.53, -8.02]$ kHz.  The final  time of the evolution is, in a), b), and c), $\approx 0.205$ ms, while in d), e), and f) $ \approx 0.411$ ms.}
\label{sup:noiseel} 
\end{figure}

In this section we show the effectiveness on noise elimination of our method. In Fig.~\ref{sup:noiseel} we have plotted the same spectrum in Fig. 2 of the main text (blue-solid), together with the individual signals of each nuclear spin (green-dashed). Blue-solid curves in Fig.~\ref{sup:noiseel}(a), (b), and (c) employ $\Omega_0 = \Omega_1 = (2\pi) \times 1$ MHz and a final sequence time of $\approx 0.205$ ms (these are the same parameters used in Fig. 2(a) of the main text). In Fig.~\ref{sup:noiseel}(d), (e), and (f) we have $\Omega_0 = (2\pi) \times 1$ MHz, $\Omega_1 = (2\pi) \times 0.5$ MHz, and a final sequence time of $\approx 0.411$ ms (as in Fig. 2(b) of the main text). Dashed-green lines  correspond to single spin signals, i.e. to the spectrum one will get if only one spin of the three included in the spin cluster is present. In Fig.~\ref{sup:noiseel}(a) and (c) we observe how we can resolve single spin signals, i.e. there is a clear overlap between the spectrum of the spin cluster (blue-solid curves) and the individual spin signals (green-dashed curves). This means that, in these cases, our method can efficiently eliminate the contribution of environment over single spin signals. However, in Fig.~\ref{sup:noiseel}(b), the spin cluster spectrum hides the presence of the single spin signal. This situation is solved in Fig.~\ref{sup:noiseel}(d), (e) and (f). Here, with the new control parameters (i.e. $\Omega_0 = (2\pi) \times 1$ MHz and $\Omega_1 = (2\pi) \times 0.5$ MHz) we can clearly resolve all the spins in the cluster and match the spin cluster signal and each individual spin signal.  Furthermore, in Fig.~\ref{sup:noiseel}(e) it is remarked with a yellow arrow a position where the value of $\langle \sigma_x\rangle$ equals to 1. This means that, for this value of $\nu$, the phase modulated sequence does not allow the NV to couple with any nuclear spin in the cluster, preserving the initial NV quantum coherence. Note also that, the latter decoupling condition can be achieved for other values of $\nu$ in each of the presented plots of Fig.~\ref{sup:noiseel} by choosing an appropriate value for $\nu$ not matching with any nuclear resonance.

\section{Power Consumption}
The average energy required to produce a certain driving can be computed as 
\begin{equation}
E=\int_{0}^{t_f}|\vec{P}|dt  
\end{equation}
where $\vec{P}=(\vec E\times \vec B)/\mu_0$ is the Poynting vector accounting for the energy per unit of area and time that an electromagnetic wave carries
being $\mu_0$ the vacuum permeability. 
The large wavelength of the microwave radiation enable us to assume the external driving as plane waves
that interact with the electronic spin throught  
$H=|\gamma_e| B_x(t) S_x$
where the magnetic field is polarised in the $x$ direction. Associated with this magnetic field there is
an electric field $\vec\nabla\times\vec E =-\partial_t \vec B$.
Without any loss of generality we can also assume that the magnetic field propagates in the $\hat y$ direction $\vec B(y,t)=B_x(t)\cos (\omega t-ky+\varphi) \hat{x}$.
After deducing the electric field related to $\vec B(y,t)$ and making use of the Poynting vector we can 
compute the energy flux of the electromagnetic wave
\begin{equation}
\label{P}
|\vec P|=\frac{2c}{\mu_0\gamma_e^2}\Omega^2(t)\tilde{c}^2(t)+\frac{2}{\mu_0\gamma_e^2k}\Omega(t)\bigg(\frac{\partial}{\partial t}\Omega(t)\bigg)\tilde{s}(t)\tilde{c}(t),
\end{equation}
with $\tilde c(t)=\cos(\omega t +\phi)$ and $\tilde s(t)=\sin(\omega t +\phi)$.
Here $c=\omega/k$, $\phi=\varphi-ky_0$ (we compute the energy flow at the NV position $y_0$), 
and  we made use of our convention $\sqrt{2}\Omega=B_x\gamma_e$ which is commented in the main text. Introducing in this expression a particular driving profile $\Omega(t)$
the energy of the electromagnetic field --i.e. the average energy, is obtained for a given sensing protocol.


\begin{thebibliography}{1}
\bibitem{Doherty13} M. W. Doherty, N. B. Manson, P. Delaney, F. Jelezko, J. Wrachtrup, and
L. C. L. Hollenberg, Phys. Rep. {\bf 528}, 1 (2013).
\bibitem{Dobrovitski13} V. V. Dobrovitski, G. D. Fuchs, A. L. Falk, C. Santori, and D. D. Awschalom,
Annu. Rev. Condens. Matter Phys {\bf 4}, 23 (2013).
\bibitem{Jelezko04} F. Jelezko, T. Gaebel, I. Popa, A. Gruber and J. Wrachtrup. Phys. Rev. Lett.
{\bf 92}, 076401 (2004).
\bibitem{Muller14} C. M\"{u}ller, X. Kong, J.-M. Cai, K. Melentijevi\'c, A. Stacey, M. Markham,
D. Twitchen, J. Isoya, S. Pezzagna, J. Meijer, J. F. Du, M. B. Plenio, B. Naydenov, L. P. McGuinness,
and F. Jelezko, Nat. Commun {\bf 5}, 4703 (2014).
\bibitem{London13} P. London, J. Scheuer, J.-M. Cai, I. Schwarz, A. Retzker, M. B. Plenio,
M. Katagiri, T. Teraji, S. Koizumi, J. Isoya, R. Fischer, L. P. McGuinness, B. Naydenov, and
F. Jelezko. Phys. Rev. Lett. {\bf 111}, 067601 (2013).
\bibitem{Fisher13} R. Fischer, C.O. Bretschneider, P. London, D. Budker, D. Gershoni, and
L. Frydman, Phys. Rev. Lett. {\bf 111}, 057601 (2013).
\bibitem{Cai13} J. M. Cai, A. Retzker, F. Jelezko, and M. B. Plenio. Nature Phys. {\bf 9}, 168 (2013).
\bibitem{Taminiau14} T. H. Taminiau, J. Cramer, T. van der Sar, V. V. Dobrovitski, and R. Hanson,
Nat. Nanotechnol. {\bf 9}, 171 (2014).
\bibitem{Waldherr14} G. Waldherr, Y. Wang, S. Zaiser, M. Jamali, T. Schulte- Herbr\"{u}ggen, H. Abe,
T. Ohshima, J. Isoya, J. F. Du, P. Neumann, and J. Wrachtrup, Nature {\bf 506}, 204 (2014).
\bibitem{Casanova15} J. Casanova, Z.-Y. Wang, J. F. Haase, and M. B. Plenio, Phys. Rev. A  {\bf 92}, 042304 (2015).
\bibitem{Cramer16} J. Cramer, N. Kalb, M. A. Rol, B. Hensen, M. S. Blok, M. Markham, D. J. Twitchen,
R. Hanson, and T. H. Taminiau, Nat. Commun. {\bf 7}, 11526 (2016).
\bibitem{Casanova16} J. Casanova, Z.-Y. Wang, and M. B. Plenio, Phys. Rev. Lett.  {\bf 117}, 130502 (2016).
\bibitem{Perlin17} M. A. Perlin, Z.-Y. Wang, J. Casanova, and M. B. Plenio, Quantum Sci. Technol. 4, 015007 (2018).
\bibitem{SchmittGS17} S. Schmitt, T. Gefen, F. M. St\"urmer, T. Unden, G. Wolff, Ch. M\"uller,
J. Scheuer, B. Naydenov, M. Markham, S. Pezzagna, J. Meijer, I. Schwarz, M. B. Plenio,
A. Retzker, L. P. McGuinness, and F. Jelezko, Science {\bf 351}, 832 (2017).
\bibitem{BossCZ17} J. M. Boss, K.S. Cujia, J. Zopes, and C. L. Degen, Science {\bf 351}, 837
(2017).
\bibitem{GlennBL18}
D. R. Glenn, D. B. Bucher, J. Lee, M. D. Lukin, H. Park and R. L. Walsworth, Nature {\bf 555},
351 (2018).
\bibitem{Wu16} Y. Wu, F. Jelezko, M. B. Plenio, and T. Weil, Angew. Chem. Int. Ed. {\bf 55}, 6586
(2016).
\bibitem{Bermudez11} A. Bermudez, F. Jelezko, M. B. Plenio and A. Retzker. Phys. Rev. Lett. {\bf 107}, 150503 (2011).
\bibitem{Cai12} J.-M. Cai, B. Naydenov, R. Pfeiffer, L. McGuinness, K. D. Jahnke, F. Jelezko,
M. B. Plenio, an A. Retzker, New J. Phys. {\bf 14}, 113023 (2012).
\bibitem{Hirose12} M. Hirose, C. D. Aiello, and P. Cappellaro. Phys. Rev. A {\bf 86}, 062320 (2012).
\bibitem{Cai13a} J. M. Cai, F. Jelezko, M. B. Plenio and A. Retzker. New J. Phys. {\bf 15}, 013020 (2013).
\bibitem{Maudsley86} A. A. Maudsley, J. Magn. Reson. {\bf 69}, 488 (1986).
\bibitem{Uhrig08} G. Uhrig, New J. Phys. {\bf 10}, 083024 (2008).
\bibitem{Pasini08} S. Pasini, T. Fischer, P. Karbach, and G. S. Uhrig, Phys. Rev. A {\bf 77}, 032315 (2008).
\bibitem{Pasini08bis} S. Pasini and G. S. Uhrig, J. Phys. A: Math. Theor. {\bf 41}, 312005 (2008).
\bibitem{Souza11} A. M. Souza, G. A. Alvarez, and D. Suter, Phys. Rev. Lett. {\bf 106}, 240501 (2011).
\bibitem{Wang11} Z.-Y. Wang and R.-B. Liu, Phys. Rev. A {\bf 83}, 022306 (2011).
\bibitem{Souza12} A. M. Souza, G. A. \'Alvarez, and D. Suter, Phil.Trans. R. Soc. A  {\bf 370},
4748 (2012).
\bibitem{Wang16} Z.-Y. Wang, J. F. Haase, J. Casanova, and M. B. Plenio, Phys. Rev. B {\bf 93}, 174104 (2016).
\bibitem{Lang17} J. E. Lang, J. Casanova, Z.-Y. Wang, M. B. Plenio, and T. S. Monteiro, Phys. Rev. Applied {\bf 7}, 054009 (2017).
\bibitem{Haase17} J. F. Haase, Z.-Y. Wang, J. Casanova, and M. B. Plenio, Phys. Rev. Lett. {\bf 121}, 050402 (2018).
\bibitem{Ryan10} C. A. Ryan, J. S. Hodges, and D. G. Cory, Phys. Rev. Lett. {\bf 105}, 200402 (2010).
\bibitem{NaydenovDH11} B. Naydenov, F. Dolde, L. T. Hall, C. Shin, H. Fedder, L. C. L. Hollenberg, F. Jelezko,
J. Wrachtrup, Phys. Rev. B {\bf 83}, 081201(R) (2011).
\bibitem{Pham12} L. M. Pham, N. Bar-Gill, C. Belthangady, D. Le Sage, P. Cappellaro, M. D. Lukin,
A. Yacoby, R. L. Walsworth. Phys. Rev. B {\bf 86}, 045214 (2012).
\bibitem{McGuinness11} L.P. McGuinness, Y. Yan, A. Stacey, D.A. Simpson, L.T. Hall, D. Maclaurin,
S. Prawer, P. Mulvaney, J. Wrachtrup, F. Caruso, R.E. Scholten, and L.C.L. Hollenberg, Nature Nanotech.
{\bf 6}, 358 (2011).
\bibitem{Cao17} Q.-Y. Cao, Z.-J. Shu, P.-C. Yang, M. Yu, M.-S. Gong, J.-Y. He, R.-F. Hu, A. Retzker,
M. B. Plenio, C. M\"{u}ller, N. Tomek, B. Naydenov, L. P. McGuinness, F. Jelezko, and J.-M. Cai, arXiv:1710.10744.
\bibitem{Levitt08} M. H. Levitt, {\em Spin dynamics: Basics of nuclear magnetic resonance}, (Wiley, 2008).
\bibitem{Gurudev07} M. V. Gurudev Dutt, L. Childress, L. Jiang, E. Togan, J. Maze, F. Jelezko, A. S. Zibrov, P. R. Hemmer, and M. D. Lukin, Science {\bf 316}, 1312 (2007).
\bibitem{Neumann10} P. Neumann, J. Beck, M. Steiner, F. Rempp, H. Fedder, P. R. Hemmer, J. Wrachtrup, and F. Jelezko, Science {\bf 329}, 542 (2010).
\bibitem{Robledo11} L. Robledo, L. Childress, H. Bernien, B. Hensen, P. F. A. Alkemade, and R. Hanson, Nature {\bf 477}, 574 (2011).
\bibitem{Sar12} T. van der Sar, Z. H. Wang, M. S. Blok, H. Bernien, T. H. Taminiau, D. M. Toyli, D. A. Lidar, D. D. Awschalom, R. Hanson, and V. V. Dobrovitski, Nature {\bf 484}, 82 (2012).
\bibitem{Liu13} G.-Q. Liu, H. C. Po, J. Du, R.-B. Liu, and X.-Y. Pan, Nat. Commun. {\bf 4}, 2254 (2013).
\bibitem{Kolkowitz12} S. Kolkowitz, Q. P. Unterreithmeier, S. D. Bennett, and M. D. Lukin, Phys. Rev. Lett. {\bf 109}, 137601 (2012).
\bibitem{Taminiau12} T. H. Taminiau, J. J. T. Wagenaar, T. van der Sar, F. Jelezko, V. V. Dobrovitski, and R. Hanson, Phys. Rev. Lett. {\bf 109}, 137602 (2012).
\bibitem{Zhao12} N. Zhao, J. Honert, B. Schmid, M. Klas, J. Isoya, M. Markham, D. Twitchen, F. Jelezko, R.-B. Liu, H. Fedder, and J. Wrachtrup, Nat. Nanotechnol. {\bf 7}, 657 (2012).
\bibitem{Hensen15} B. Hensen, H. Bernien, A. E. Dr\'eau, A. Reiserer, N. Kalb, M. S. Blok, J. Ruitenberg, R. F. L. Vermeulen,
R. N. Schouten, C. Abell\'an, W. Amaya, V. Pruneri, M. W. Mitchell, M. Markham, D. J. Twitchen, D. Elkouss, S. Wehner, T. H. Taminiau,
and R. Hanson, Nature {\bf 526}, 682 (2015).
\bibitem{Lovchinsky16} I. Lovchinsky, A. O. Sushkov, E. Urbach, N. P. de Leon, S. Choi, K. De Greve, R. Evans, R. Gertner, E. Bersin, C. Müller, L. McGuinness, F. Jelezko, R. L. Walsworth, H. Park, M. D. Lukin, Science
{\bf 351}, 836 (2016).
\bibitem{Abobeih18} M. H. Abobeih, J. Cramer, M. A. Bakker, N. Kalb, M. Markham, D. J. Twitchen, and T. H. Taminiau, E-Print arXiv:1801.01196.
\bibitem{Hartmann62} S. Hartmann and E. Hahn, Phys. Rev. {\bf 128}, 2042 (1962).
\bibitem{Casanova18} J. Casanova, Z.-Y. Wang, I. Schwartz, and M. B. Plenio, Phys. Rev. Applied {\bf 10}, 044072 (2018).
\bibitem{Maze08} J. R. Maze, J. M. Taylor, and M. D. Lukin, Phys. Rev. B {\bf 78}, 094303 (2008).
\bibitem{Supmat} See Supplemental Material for further details.
\bibitem{Laraoui11} A. Laraoui and C. A. Meriles,  Phys. Rev. B {\bf 84}, 161403(R) (2011).
\bibitem{Naydenovp} B. Naydenov, private communication.
\bibitem{Childress06} L. Childress, M. V. Gurudev Dutt1, J. M. Taylor A. S. Zibrov, F. Jelezko, J. Wrachtrup, P. R. Hemmer, and M. D. Lukin, Science {\bf 314}, 281 (2006).
\bibitem{Uhlenbeck30} G. E. Uhlenbeck, and L. Ornstein Phys. Rev. {\bf 36}, 36823 (1930).
\bibitem{Cohen17} I. Cohen, N. Aharon, and A. Retzker, Fortschr. Phys. {\bf 65}, 1600071 (2017)

\end{thebibliography}

\begin{thebibliography}{99} 
\end{thebibliography}
\end{document}